\documentstyle[twocolumn,seceq,epsfig]{jpsj}

\def\runtitle{CTMRG Method Applied to the Ising Model on the Hyperbolic Plane}
\def\runauthor{Kouji {\sc Ueda}, Roman {\sc Krcmar}, Andrej {\sc Gendiar}, 
 and Tomotoshi {\sc Nishino}}

\title{Corner Transfer Matrix Renormalization Group Method Applied to the Ising Model 
on the Hyperbolic Plane}

\author
{Kouji {\sc Ueda}$^{1)}$, Roman {\sc Krcmar}$^{2)}$, Andrej {\sc Gendiar}$^{2)}$,  
 and Tomotoshi {\sc Nishino}$^{1)}$}
\inst
{$^1$Department of Physics, Graduate School of Science, 
Kobe University, Kobe 657-8501, Japan\\
$^2$Institute of Electrical Engineering, Slovak Academy of Sciences,
SK-841 04 Bratislava, Slovakia
}

\abst
{Critical behavior of the Ising model  is investigated at the center of large scale
finite size systems, where the lattice is represented as 
the tiling of pentagons. The system is 
on the hyperbolic plane, and the recursive structure of the lattice makes it possible 
to apply the corner transfer matrix renormalization group method. From the calculated 
nearest neighbor spin correlation function and the spontaneous magnetization, 
it is concluded that the phase transition of this model is mean-field like. One parameter
deformation of the corner Hamiltonian on the hyperbolic plane is discussed.}

\kword{CTMRG, Ising Model, Scaling, Curvature}

\begin{document}
\sloppy
\maketitle

\section{Introduction}

Baxter's method of corner transfer matrix (CTM) has been known as one of the
representative tool for analytical stydy of statistical models in two 
dimension (2D).~\cite{Bax1,Bax2,Bax3}
The method is also of use for numerical calculations of one point functions, 
such as the local energy and the magnetization.~\cite{Bax3} This
numerical application is a kind of numerical renormalization group (RG) method, where
the block spin transformation is obtained from the diagonalization of CTMs. 
Such a RG scheme has many aspects in common with the density matrix 
renormalization group (DMRG) method,~\cite{White1,White2,Dresden,Sch} 
expecially when the method is applied to 2D classical lattice models.~\cite{Nishino}

Introducing the flexibility in the system extension process of the DMRG method to
the Baxter's method of CTM, the authors developed the 
corner transfer matrix renormalization group (CTMRG) 
method.~\cite{CTMRG1,CTMRG2,CTMRG3,CTMRG4}
In this article we report a modification of the CTMRG method, 
for the purpose of applying the method to 
classical lattice models on the hyperbolic plane. 
Using the recursive structure of the lattice, we obtain 
one point functions at the center of sufficiently large finite size systems.

Quite recently Hasegawa, Sakaniwa, and Shima reported deviations of 
critical indices of the Ising model on the hyperbolic plane 
from the well known Ising universality classes in two dimension.~\cite{Shima1,Shima2} 
They predicted that phase transition of such systems would be mean-field like. 
To confirm their prediction, in the next section we consider the Ising model 
on a lattice, which is represented as the tiling of pentagons. The necessary
modification of the CTMRG method on this lattice is explained in \S 3.
We calculate the nearest neighbor spin correlation function and the 
spontaneous magnetization at the center of large scale finite size systems. 
Critical indices for these one point functions
are studied in \S 4, and we confirme the mean-field like properties of the
phase transition. Conclusions are summarized in the last section.
We discuss a possible deformation of corner Hamiltonian in the 
hyperbolic plane.

\section{Ising Model on the Tiling of Pentagons}

Let us consider the hyperbolic plane, which is the two 
dimensional surface with constant negative curvature. Figure 1 shows a 
part of a sufficiently large regular lattice on the plane, where 
the lattice is constructed as the tiling of pentagons.~\cite{URL}
All the arcs are geodesics, which divides the lattice into two 
parts of similar structure. Each lattice point represented by an open circle is the
crossing points of two geodesics.

\begin{figure}
\epsfxsize=50mm 
\centerline{\epsffile{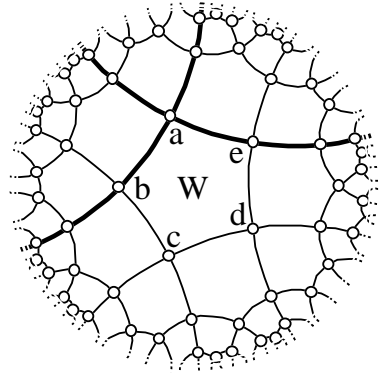}}
\caption{Ising model on a regular lattice on the hyperbolic plane. 
Open circles represent Ising spins on the lattice point, and 
 $W( \sigma_a^{~}, \sigma_b^{~}, \sigma_c^{~}, \sigma_d^{~}, \sigma_e^{~} )$ 
is the local Boltzmann weight defined in Eq.~(2.2).
Two geodesics drawn by thick arcs divide the system into four quadrants, which
are called as the corner. }
\label{fig:1}
\end{figure}

We investigate the ferromagnetic Ising model on this lattice. 
The Hamiltonian of the system is defined as the sum of nearest 
neighbor Ising interactions
\begin{equation}
H = - J \sum_{\langle i, j \rangle}^{~} \sigma_i^{~} \sigma_j^{~} \, ,
\end{equation}
where $\langle i, j \rangle$ represents pair of neighboring sites,
and $\sigma_i^{~} = \pm 1$ and  $\sigma_j^{~} = \pm 1$ are the Ising 
spins on the lattice points. Throughout this article we assume the absence
of external magnetic field.
For the latter conveniences, we represent the system as the `interaction round a 
face (IRF)' model. The local Boltzmann weight for each face of pentagonal 
shape --- the IRF weight --- is given by
\begin{eqnarray}
&&W( \sigma_a^{~}, \sigma_b^{~}, \sigma_c^{~}, \sigma_d^{~}, \sigma_e^{~} ) 
\\
&&= 
\exp\left\{ - \frac{\beta J}{2} \left( 
\sigma_a^{~} \sigma_b^{~} + 
\sigma_b^{~} \sigma_c^{~} + 
\sigma_c^{~} \sigma_d^{~} + 
\sigma_d^{~} \sigma_e^{~} + 
\sigma_e^{~} \sigma_a^{~}
\right)\right\} \nonumber
\end{eqnarray}
where $\beta = 1 / k_{\rm B}^{~} T$ is the inverse temperature, and 
$\sigma_a^{~}$, $\sigma_b^{~}$, $\sigma_c^{~}$, 
$\sigma_d^{~}$, and $\sigma_e^{~}$ are the spin variables 
around the face as shwon in Fig.~1. 

The partition function of a finite size system (with sufficiently large diameter)
is formally written as the configuration sum of the Boltzmann weight 
of the whole system
\begin{equation}
Z = 
\sum_{\hbox{\tiny\rm all the} \atop \hbox{\tiny\rm spins}}^{~} 
\prod_{\hbox{\tiny\rm all the} \atop \hbox{\tiny\rm faces}}^{~} \, 
W \, ,
\end{equation}
where we are interested in the thermodynamic limit of this system.
Note that it is rather hard to investigate the system by use of the Monte 
Carlo simulations, since the number of sites contained in a cluster blows up 
exponentially with respect to its diameter. As a complemental numerical
tool, we employ the CTMRG method.

\section{Corner Transfer Matrix Renormalization Group Method}

The two geodesics shown by thick arcs in Fig.~1 divide the system into four 
parts, which are called as corners.~\cite{Bax3} 
Figure 2 shows the structure of a corner. We label the spins on a cut as 
$\{ \sigma_1^{~}, \sigma_2^{~}, \sigma_3^{~}, \ldots \}$, and
those on another cut as
$\{ \sigma'_1, \sigma'_2, \sigma'_3, \ldots \}$, 
where $\sigma_1^{~}$ is equivalent to $\sigma'_1$.
The corner transfer matrix is the Boltzmann weight with respect to
a corner, which is calculated as a partial sum of the product of IRF 
weights in the corner
\begin{eqnarray}
&&C( \sigma'_1, \sigma'_2, \sigma'_3, \ldots | \, 
\sigma_1^{~}, \sigma_2^{~}, \sigma_3^{~}, \ldots ) \nonumber\\
&& ~~~~~~~ =
\sum_{\hbox{\tiny\rm spins inside} \atop \hbox{\tiny\rm the quadrant}}^{~} 
\prod_{\hbox{\tiny\rm faces in} \atop \hbox{\tiny\rm the quadrant}}^{~} \, 
W \, .
\end{eqnarray}
The configuration sum is taken over spins `inside' the corner, leaving
those spins on the cuts. Conventionally the matrix
$C$ is interpreted as block diagonal with respect to $\sigma_1^{~}$ and
$\sigma'_1$, and the element $C( \sigma'_1, \sigma'_2, \sigma'_3, \ldots | \, 
\sigma_1^{~}, \sigma_2^{~}, \sigma_3^{~}, \ldots )$ for
those cases $\sigma_1^{~} \ne \sigma'_1$ is set to zero. The CTM thus defined
is symmetric in the case of the pentagonal lattice under consideration.

\begin{figure}
\epsfxsize=50mm
\centerline{\epsffile{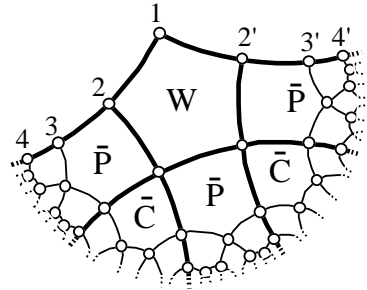}}
\caption{Recursive structure of a corner $C$.}
\label{fig:2}
\end{figure}

As shown in Fig.~2 a corner has the structure where three parts 
labeled by $\bar P$ and two parts labeled by $\bar C$ are joined to 
a face $W$. The `fusion' relation can be represented by a formal 
equation~\cite{CTMRG1,CTMRG2}
\begin{equation}
C = W \cdot \bar P \bar C \bar P \bar C \bar P \bar C \, .
\end{equation}
Note that $\bar C$ is a corner of smaller size. 

For convenience, let us observe the structure of the part of the system shown
in Fig.~3, where two $\bar P$, and $\bar C$ are joined to $W$.
Labeling the shown part by $P$, we can formally write the fusion relation in the 
same manner
\begin{equation}
P = W \cdot \bar P \bar C \bar P \, .
\end{equation}
For a conventional reason we call $P$ as the `half-row', although $P$ is not
a row on the hyperbolic plane.
It is easily understood that  $\bar P$ is a half-row of smaller size.
We have thus obtained recursive structure of the corner $C$ and the half-row $P$.

\begin{figure}
\epsfxsize=50mm
\centerline{\epsffile{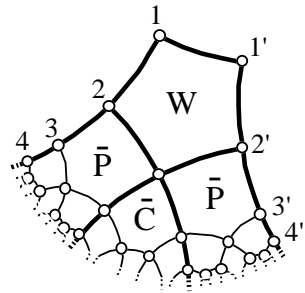}}
\caption{Recursive structure of the `half-raw' $P$.}
\label{fig:3}
\end{figure}

As we have defined CTM for a corner, let us express the Boltzmann
weight with respect to the half-row
\begin{eqnarray}
&&P( \sigma'_1, \sigma'_2, \sigma'_3, \ldots | \, 
\sigma_1^{~}, \sigma_2^{~}, \sigma_3^{~}, \ldots ) \nonumber\\
&& ~~~~~~~ =
\sum_{\hbox{\tiny\rm spins inside} \atop \hbox{\tiny\rm the half-row}}^{~} 
\prod_{\hbox{\tiny\rm faces in} \atop \hbox{\tiny\rm the half-row}}^{~} \, 
W 
\end{eqnarray}
in the matrix form, where the positions of spins
$\{ \sigma_1^{~}, \sigma_2^{~}, \sigma_3^{~}, \ldots \}$ and
$\{ \sigma'_1, \sigma'_2, \sigma'_3, \ldots \}$ are shown in Fig.~3.
We call the weight in the matrix form as the half-row transfer matrix (HRTM).

The 4-th power of the CTM 
\begin{equation}
\rho = C^4_{~} 
\end{equation}
is a kind of density matrix, since its trace gives the partition function
\begin{equation}
Z = {\rm Tr} \, \rho = {\rm Tr} \, C^4_{~} 
\end{equation}
of a finite size cluster that consists of four corners.
The matrix dimension of CTM, and also that of the HRTM, 
increases exponentially with respect to the system size. 
In order to obtain $Z$ numerically up to sufficiently large systems,
we introduce the block spin transformation that is created from the 
diagonalization of the density matrix $\rho$.~\cite{White1,White2} 

Instead of directly diagonalizing $\rho$ in Eq.~(3.5), we first create its contraction 
\begin{eqnarray}
&&\rho'( \sigma'_2, \sigma'_3 \ldots | \, \sigma_2^{~}, \sigma_3^{~}, \ldots )
\nonumber\\
&&~~~~ = \sum_{\sigma'_1 = \sigma_1^{~} = \pm 1}^{~}
\rho( \sigma'_1, \sigma'_2, \sigma'_3 \ldots | \, 
\sigma_1^{~}, \sigma_2^{~}, \sigma_3^{~}, \ldots ) 
\end{eqnarray}
and then diagonalize it
\begin{eqnarray}
&&\rho'( \sigma'_2, \sigma'_3 \ldots | \, \sigma_2^{~}, \sigma_3^{~}, \ldots )
\nonumber\\
&&~~~~ = \sum_{\xi}^{~}
A( \sigma'_2, \sigma'_3 \ldots | \, \xi ) \, \lambda_{\xi}^{~} \,
A( \sigma_2^{~}, \sigma_3^{~}, \ldots | \, \xi ) \, ,
\end{eqnarray}
where the eigenvalue $\lambda_{\xi}^{~}$ is non-negative. Following the
convention in DMRG, we assume the decreasing order for $\lambda_{\xi}^{~}$.
The orthogonal matrix $A( \sigma_2^{~}, \sigma_3^{~}, \ldots | \, \xi )$ 
represents the block spin transformation 
from the `row-spin' $\{ \sigma_2^{~}, \sigma_3^{~}, \ldots \}$ to
the effective spin variable $\xi$. We keep $m$ numbers of representative
states, which correspond to major eigenvalues, for the block spin variable $\xi$.
Applying the matrix $A$ to the CTM and the HRTM,  
we obtain `renormalized matrices' of $2m$-dimension
\begin{eqnarray}
C( \sigma'_1, \sigma'_2,  \ldots | \, 
\sigma_1^{~}, \sigma_2^{~}, \ldots ) &\rightarrow&
C( \sigma', \xi'_{~} | \, \sigma, \xi )  \nonumber\\
P( \sigma'_1, \sigma'_2, \ldots | \, 
\sigma_1^{~}, \sigma_2^{~}, \ldots ) &\rightarrow&
P( \sigma', \xi'_{~} | \, \sigma, \xi ) \, ,
\end{eqnarray}
where we have dropped the indices from $\sigma_1^{~}$ and
$\sigma'_1$.~\cite{CTMRG1,CTMRG2}

Combining the recursive structures in Eqs.~(3.2) and (3.3), and the
renormalization scheme in Eqs.~(3.7)-(3.9), we can obtain the 
CTM and HRTM in the renormalized form 
for arbitrary system size by way of successive extension of
the system.  

\begin{figure}
\epsfxsize=75mm 
\centerline{\epsffile{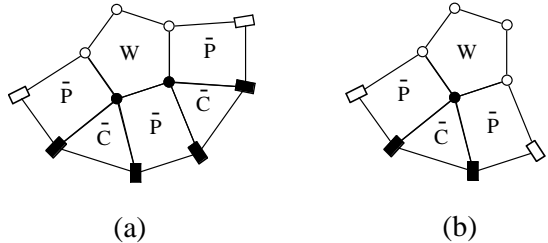}}
\caption{Extension pocess of (a) CTM and (b) HRTM in the renormalized expression.}
\label{fig:4}
\end{figure}

Suppose that we have
$C( \sigma', \xi'_{~} | \, \sigma, \xi )$ and
$P( \sigma', \xi'_{~} | \, \sigma, \xi )$ for a finite size cluster.
In order to explain the extension process, let us rewrite these matrices as
${\bar C}( s'_{~}, \zeta'_{~} | \, s, \zeta )$ and
${\bar P}( s'_{~}, \zeta'_{~} | \, s, \zeta )$. 
\vskip 2mm
\begin{itemize}
\item[(1)] Substitute  ${\bar C}( s'_{~}, \zeta'_{~} | \, s, \zeta )$ and
${\bar P}( s'_{~}, \zeta'_{~} | \, s, \zeta )$ into the fusion process in Eqs.~(3.2) and (3.3).
Figure 4 shows these fusion processes among $W$, $\bar C$, and $\bar P$, where
rectangles correspond to the block spin variables. The spin variables that are
contracted out are shown by black marks. As a result, we obtain the extended CTM 
$C( \sigma'_{~}, s'_{~}, \zeta'_{~} | \, \sigma, s, \zeta )$ and the extended HRTM
$P( \sigma'_{~}, s'_{~}, \zeta'_{~} | \, \sigma, s, \zeta )$.
\item[(2)] From the extended CTM $C( \sigma'_{~}, s'_{~}, \zeta'_{~} | \, \sigma, s, \zeta )$
obtain the density matrix $\rho( \sigma'_{~}, s'_{~}, \zeta'_{~} | \, \sigma, s, \zeta )$ by
Eq.~(3.5). Contracting out the spin at the center, obtain $\rho'_{~}( s'_{~}, \zeta'_{~} | \, s, \zeta )$
as Eq.~(3.7), and diagonalizing it to obtain the block spin transformation matrix 
$A( s, \zeta | \, \xi )$ from Eq.~(3.8).
\item[(3)] Applying $A( s, \zeta | \, \xi )$ to both
$C( \sigma'_{~}, s'_{~}, \zeta'_{~} | \, \sigma, s, \zeta )$ and
$P( \sigma'_{~}, s'_{~}, \zeta'_{~} | \, \sigma, s, \zeta )$, 
obtain the extended CTM
$C( \sigma'_{~}, \xi'_{~} | \, \sigma, \xi )$ 
in the initial form, and the same for HRTM to obtain
$P( \sigma'_{~}, \xi'_{~} | \, \sigma, \xi )$.
\item[(4)] return to the first step.
\end{itemize}
\vskip 2mm
The system size, which is the length of the longest geodesics in the 
system, increases by 2 for each iteration.~\cite{boundary}
In order to start the above extension process,
we set the initial condition
\begin{equation}
{\bar C}( \sigma'_{~} | \, \sigma ) = {\bar P}( \sigma'_{~} | \, \sigma ) =
\delta( \sigma'_{~} | \, 1 ) \, \delta( \sigma | \, 1 ) 
\end{equation}
that represents ferromagnetic boundary,
where $\delta( a | \, b ) = \delta_{a, b}^{~}$ is the Cronecker's delta.

During the iteration we can obtain one point functions at the center of the system.
For example, the spontaneous magnetization is calculated as
\begin{equation}
\langle \sigma \rangle = \frac{{\rm Tr} \, \sigma \rho}{{\rm Tr} \, \rho} 
=
\frac{\displaystyle \sum_{\sigma, s, \zeta}^{~} \sigma
\rho( \sigma, s, \zeta | \, \sigma, s, \zeta )
}{\displaystyle \sum_{\sigma, s, \zeta}^{~} \rho( \sigma, s, \zeta | \, \sigma, s, \zeta )
} \, .
\end{equation}
In the same manner we obtain the nearest neighbor spin correlation function 
$\langle \sigma s \rangle$. It should be noted that one point functions thus 
calculated at the center do not always represent the averaged property of the whole system
even in the thermodynamic limit, since the area near the boundary has non-negligible
weight in the hyperbolic plane.

\section{Numerical Result Compared With the Bethe Approximation}

Let us calculate the spontaneous magnetization $\langle \sigma \rangle$, 
and spin correlation function $\langle \sigma s \rangle$ 
for the nearest spin pair. We regard the Ising interaction strength $J$ as the 
energy unit, and use the temperature where the Boltzmann constant 
$k_{\rm B}^{~}$ is equal to unity. Most of the numerical calculations are performed
keeping $m = 40$ states. The dumping of the density matrix eigenvalues is
very fast, and actually the calculated results with $m = 10$ do not differ
from those obtained with $m = 40$ even at the critical temperature $T_{\rm C}^{~}$.
The iteration number required for the numerical convergence is
at most 400000 for the calculated data points.

\begin{figure}
\epsfxsize=75mm 
\centerline{\epsffile{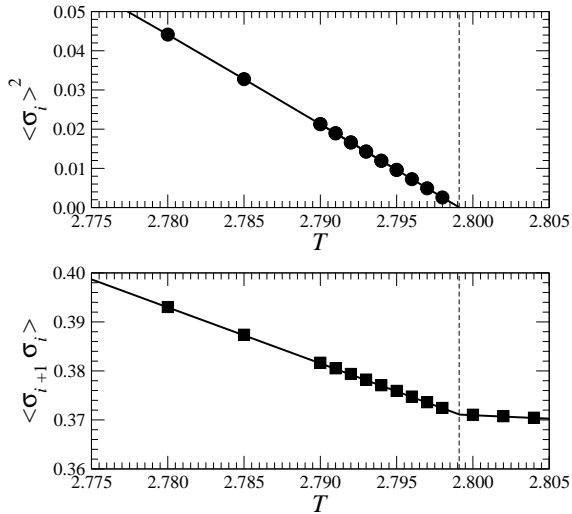}}
\caption{Square of the spontaneous magnetization 
$\langle \sigma \rangle^2_{~}$ (upper) and the nearest 
neighbor spin correlation function 
$\langle \sigma s \rangle$ (lower) with respect to the temperature $T$.}
\label{fig:5}
\end{figure}

Figure 5 shows the calculated results. The square of the spontaneous
magnetization $\langle \sigma \rangle^2_{~}$ is a linear function of 
temperature in the neighborhood of $T_{\rm C}^{~}$.
From the behavior we estimate the transition temperature $T_{\rm C}^{~} = 2.799$.
The nearest neighbor spin correlation function $\langle \sigma s \rangle$
has a kink at $T_{\rm C}^{~}$, and is linear in $T$ around there. These
calculated results support the existence of mean-field like transition, that is
subject to the critical indices $\beta = 1 / 2$ and $\alpha = 0$. We thus
confirmed the prediction by Hasegawa, Sakaniwa, and Shima.~\cite{Shima1,Shima2}


Compared with the transition temperature of the
square lattice Ising model $T_{\rm C}^{\rm Square} = 2.269$, 
the calculated $T_{\rm C}^{~}$ is fairly higher and
is close to the transition temperature calculated from the  Bethe 
approximation $T_{\rm C}^{\rm Bethe} = 2.885$.~\cite{Bethe,same}
The result suggest that neglection of the `loop back effect' is not
so conspicuous in the hyperbolic plane. 


\section{Conclusion and discussion}

We have calculated the spontaneous magnetization and the nearest neighbor
spin correlation function of the Ising model on a pentagonal lattice on the 
hyperbolic plane. The numerical algorithm of the CTMRG method is modified 
for this purpose. The calculated critical temperature is $T_{\rm C}^{~} = 2.799$,
and we observe the mean-field like phase transition. 

The modified CTMRG method we have developped is applicable to regular lattices that
consists of geodesics on the hyperbolic plane. For those lattices that does
not contain geodesics, one has to either treating asymmetric density matrix
or to draw geodesics by use of transformation such as duality transformation 
and the star-triangle relation. Generalization of the modified CTMRG method 
to the vertex model is straight forward. It may be interesting to classify 
ordered states of eight-vertex model on a variety of regular lattices in the 
hyperbolic plane.

An interest is in the eigenvalue structure of the density matrix at the transition 
temperature. Its analytic form is not well defined in the thermodynamic limit 
of classical lattice models on the flat 2D plane.~\cite{Peschel,Okunishi1} The 
rapid eigenvalue dumping observed on
the hyperbolic plane suggests that there would be a way of regularizing the CTM
at the criticality.
The classical-quantum correspondence from such a view point is worth considering.
Formally speaking the corner transfer matrix $C$ can be written as the exponential of the 
corner Hamiltonian $H_{\rm C}^{~}$. When the lattice is on the flat plane, the simplest
example of the corner Hamiltonian is written in the sum of local operators
\begin{eqnarray}
H_{\rm C}^{~} 
&=& 
h( \sigma_1^{~}, \sigma_2^{~} ) + 
2 h( \sigma_2^{~}, \sigma_3^{~} ) +
3 h( \sigma_3^{~}, \sigma_4^{~} ) + \ldots \nonumber\\
&=&
h_1^{~} + 2 h_2^{~} + 3 h_3^{~} + 4 h_4^{~} + \ldots \, ,
\end{eqnarray}
where $h_i^{~} = h( \sigma_i^{~}, \sigma_{i+1}^{~} )$ is the local hamiltonian that acs
between neighboring sites. A possible one parameter deformation of the 
corner Hamiltonian to the hyperbolic geometry may given by
\begin{equation}
H_{\rm C}^{~}( \Lambda ) = h_1^{~} + 
\frac{\sinh 2 \Lambda}{\sinh \Lambda} h_2^{~} +
\frac{\sinh 3 \Lambda}{\sinh \Lambda} h_3^{~} + \ldots \, ,
\end{equation}
which is reduced to $H_{\rm C}^{~}$ in Eq.~(5.1) in the limit $\Lambda \rightarrow 0$.
The deformed Hamiltonian satisfies the recursive structure
\begin{eqnarray}
H_{\rm C}^{~}( \Lambda ) 
&=& 
\cosh \Lambda \left( h_2^{~} +
\frac{\sinh 2 \Lambda}{\sinh \Lambda} h_3^{~} + 
\frac{\sinh 3 \Lambda}{\sinh \Lambda} h_4^{~} + \ldots \right) \nonumber\\
&+&
h_1^{~} + \cosh \Lambda \, h_2^{~} + \cosh 2 \Lambda \, h_3^{~} + \ldots
\end{eqnarray}
discussed in Okunishi's RG scheme on the corner Hamiltonian.~\cite{Okunishi2} 
Such a deformation has similar regularization effect proposed by Okunishi quite 
recently.~\cite{Okunishi3}

T.~N. and A.~G is partially supported by a Grant-in-Aid for Scientific Research from the Ministry of 
Education, Science, Sports and Culture. T.~N. thank to Okunishi for valuable discussions.


\begin{thebibliography}{99}
\bibitem{boundary} Precisely speaking, such a way of system extension does not
guarantee the cluster shape where every spins at the boundary is equally distant
from the center.
\bibitem{same} Since the configuration number of
the lattice is $z = 4$, the mean-field approximation applied to the 
Ising model on the pentagonal lattice under consideration gives the 
same transition temperature as the approximation applied to the square 
lattice Ising model. Such an equivalence holds up to the Bethe approximation.
\end{thebibliography}
\end{document}